\documentclass[twocolumn,pra,aps,superscriptaddress,floatfix]{revtex4}
\usepackage{amssymb}
\usepackage{amsfonts}
\usepackage{amsmath}
\usepackage{graphicx}

\begin{document}

\title{Generalized Loschmidt echoes associated with operational quantum
non-Markovianity}
\date{\today }
\author{Cecilia Cormick}
\affiliation{Instituto de F\'isica de la Facultad de Ingenier\'ia, Universidad de la
Rep\'ublica, Julio Herrera y Reissig 565, Montevideo, Uruguay}
\author{Adri\'{a}n A. Budini}
\affiliation{Consejo Nacional de Investigaciones Cient\'{\i}ficas y T\'{e}cnicas
(CONICET), Centro At\'{o}mico Bariloche, Avenida E. Bustillo Km 9.5, (8400)
Bariloche, Argentina, and Universidad Tecnol\'{o}gica Nacional (UTN-FRC),
Fanny Newbery 111, (8400) Bariloche, Argentina}

\begin{abstract}
Non-Markovianity can be characterized by performing a series of successive
measurements and analyzing departures of the corresponding outcome
statistics from a Markovian probabilistic structure. Considering a system
coupled with its environment via a dephasing interaction, we show that joint
outcome probabilities can be written in terms of a set of two-time
environment correlations. Their definition involves forward and backward
propagators with different Hamiltonians, associated with a recently
introduced generalization of standard Loschmidt echoes [Cormick and Budini,
Phys. Lett. A \textbf{593}, 132011 (2026)]. This result establishes a solid
connection between quantum non-Markovianity defined in an operational
(measurement-based) way and complex quantum dynamics studied through their
sensitivity to dynamical perturbations. We find conditions that guarantee a
Markovian (system) behavior and also determine how the generalized echoes
can identify information exchanges between the system and its environment.
We illustrate our ideas considering examples of spin environments that
realize these different dynamical regimes.
\end{abstract}

\maketitle

\section{Introduction}

The Loschmidt echo (LE) is a conceptual and theoretical physical tool that
emerged in the context of experimental nuclear magnetic resonance~\cite%
{Ernst_1992, Usaj_1995, Pastawski_1998}. From a formal point of view it can
be defined as the overlap between two quantum states of a complex system.
Starting from a given initial condition, one of them corresponds to the
standard forward time-evolution while the other one includes a perturbation
from the original complex dynamics. Equivalently, this overlap can be
interpreted as a revival probability after a forward evolution followed by a
time-reversal protocol containing small unavoidable departures from the
exact time-reversed Hamiltonian. The relevance of the LE as a theoretical
tool comes from significant information that is encoded in this overlap. In
fact, its characterization involves areas such as quantum and semiclassical
chaos, random matrix theory, solid state physics, many body systems, and
quantum complexity, just to name a few~\cite{jalabert,beenaker,dalvit,
Wisniacki_2003,fernando,prosen,quan,vidal,
Paz_2008,calvo,danieli,dente,Pastawski_2013_PRL, Pastawski_2013_PRB,
zangara,yan,karina,hase,sanchez,kumar}.

The experimental measurement of LEs generally involves a quantum system of
small dimensionality that in turn is coupled to (or is part of) a complex
one characterized by a much higher dimensionality. Therefore, the complex
system can be read as the environment of the small one. Although the
original perspectives are quite different, this splitting also puts LEs in
the frame of the theory of open quantum systems~\cite{breuerbook,vega}. In
this latter context, the interest is to characterize the dynamical
properties of the small system as a function of the properties of the
environment and the coupling to it.

One of the cornerstones of the theory of open quantum systems is the
Born-Markov approximation, which permits the description of the system
dynamics in terms of Lindblad equations. Thus, the time evolution is
approximately governed by quantum master equations that are local in time
and fulfill a semigroup property~\cite{breuerbook}. Departures from this
limit have attracted much interest in the last decades~\cite{vega}.
Furthermore, the definition and quantification of non-Markovianity in the
theory of open quantum systems led, broadly speaking, to two complementary
research lines. In \textit{non-operational} approaches to quantum
non-Markovianity \cite%
{BreuerReview,plenioReview,BreuerFirst,cirac,rivas,breuerDecayTLS,fisher,fidelity,dario,mutual,geometrical,DarioSabrina,brasil,canonicalCresser,Acin,poland,BreuerExp,breuerDrift,urrego,khurana,sun,chile,pan}%
, memory effects are determined and quantified from departures of the system
propagator with respect to a Lindblad equation. Instead, in \textit{%
operational} approaches the relevant information, in analogy with classical
stochastic systems, is given by the outcome statistics associated with a set
of (projective) measurement processes performed explicitly\ over the system
of interest~\cite{modi, Modi_2018_PRA, pollock, Modi_2019_PRA,
goan,budiniCPF, Budini_2019_PRA, budiniChina,budiniBrasil,bonifacio,han,555}.

Given the advancements in both LE techniques and definitions of quantum
non-Markovianity, points of contact between both areas have been explored in
the literature. For example, in Ref.~\cite{tana} it was shown that LEs can
be put in one-to-one correspondence with different definitions of
(non-operational) quantum non-Markovianity. The main idea is to consider a
small system coupled to a complex one through a dephasing interaction.
Non-monotonic decay of the system coherence, in particular through coherence
revivals, can be read as an environment-to-system backflow of information
that signals the presence of memory effects. Given the dephasing nature of
the interaction, the coherence decay has the structure of a LE. Hence,
fingerprints of memory effects are encoded in the time decay of the LE.

The main purpose of this contribution is to establish a robust and
well-formulated relation between LEs and quantum non-Markovianity, defined
in an operational---measurement-based---context. Since projective
measurements are not a basic ingredient of nuclear magnetic resonance
experiments, this connection is not evident. Considering a dephasing
system-environment interaction and introducing a minimal scheme of three
projective measurements, we show that outcome probabilities can be expressed
in terms of a recent generalization of the Loschmidt echoes (GLEs)~\cite%
{PLA_2026}. The definition of these quantities involves the same forward and
backward propagators that appear in the standard LE, but now including
combinations of two forward and two backward evolution steps. After
establishing the relation between outcome probabilities and GLEs, we also
find conditions on the echoes that lead to (system) Markovianity and
complementarily to system-environment information exchanges~\cite{BIF,
Budini_2022_Entropy, CBE}. In addition, these concepts are illustrated in
detail through the example of an environment corresponding to an integrable
spin chain that leads to different kinds of perturbations of the system.

The manuscript is organized as follows. In Sec. II we present the definition
of the GLEs and the underlying system-environment model that grounds our
theoretical formulation. In Sec. III we explicitly show the relation between
operational quantum non-Markovianity and GLEs, analyzing also the conditions
that correspond to the Markovian regime and system-environment information
exchanges. In Sec. IV we explicitly characterize the GLEs for a spin
environment with a perturbation that modifies both eigenvalues and
eigenstates of the environment. In Sec. V we summarize our conclusions.
Additional supporting calculations are provided in the Appendix.

\section{Generalized Loschmidt echoes}

The main goal of the following sections is to demonstrate that the set of
GLEs characterizes the non-Markovian properties of an open quantum system
when memory effects are described through an operational (measurement-based)
approach. For completeness, in this section we review the GLEs introduced in~%
\cite{PLA_2026} and also establish the underlying microscopic model as well
as the resulting evolution of the system density matrix.

\subsection{Definitions}

The LE corresponds to the overlap between two states of a complex quantum
system. Each state is evolved with a slightly different Hamiltonian.
Denoting the corresponding propagators as $\mathbb{G}_{t_{f},t_{i}}^{(s)}$
with $s=\pm ,$ where $t_{i}$\ and $t_{f}$ are initial and final times
respectively, the LE here is defined as%
\begin{equation}
\mathrm{E}(\tau )\equiv \langle \mathcal{B}|\mathbb{G}_{\tau ,0}^{(-)\dag }%
\mathbb{G}_{\tau ,0}^{(+)}|\mathcal{B}\rangle .  \label{StandardLE}
\end{equation}%
$|\mathcal{B}\rangle $ is the initial state of the complex system. The
product $\mathbb{G}_{\tau ,0}^{(-)\dag }\mathbb{G}_{\tau ,0}^{(+)}$\ is
usually called the echo operator or echo propagator. We remark that $\mathrm{%
E}(\tau )$ is usually termed the LE amplitude while its absolute squared
value, $|\mathrm{E}(\tau )|^{2}$, is termed the LE. Since phases will be
relevant for our analysis, in the present contribution for brevity we refer
to $\mathrm{E}(\tau )$ as the LE.

The expression for $\mathrm{E}(\tau )$ only involves a time interval with
two different propagators. The GLEs~\cite{PLA_2026}, which also emerge in
the analysis developed in the next sections, are defined as%
\begin{equation}
\mathrm{E}_{\tilde{s}s}(t,\tau )\equiv \langle \mathcal{B}|\mathbb{G}%
_{t,0}^{(\tilde{s})\dag }\mathbb{G}_{t+\tau ,t}^{(-)\dag }\mathbb{G}_{t+\tau
,t}^{(+)}\mathbb{G}_{t,0}^{(s)}|\mathcal{B}\rangle ,  \label{GeneralizedLE}
\end{equation}%
where $\tilde{s}=\pm $\ and $s=\pm .$ This definition involves four
propagators with two different time intervals, $\tau >0,$ and $t>0.$ The
intermediate product $\mathbb{G}_{t+\tau ,t}^{(-)\dag }\mathbb{G}_{t+\tau
,t}^{(+)}$ is the standard echo operator shifted in time, which in turn is
contracted with the (evolved) states $\mathbb{G}_{t,0}^{(s)}|\mathcal{B}%
\rangle $ and $\langle \mathcal{B}|\mathbb{G}_{t,0}^{(\tilde{s})\dag }.$
Using the divisibility of the (unitary) propagators, the four GLEs
explicitly read 
\begin{subequations}
\label{ExplicitGLE}
\begin{eqnarray}
\mathrm{E}_{++}(t,\tau ) &=&\langle \mathcal{B}|\mathbb{G}_{t,0}^{(+)\dag }%
\mathbb{G}_{t+\tau ,t}^{(-)\dag }\mathbb{G}_{t+\tau ,0}^{(+)}|\mathcal{B}%
\rangle , \\
\mathrm{E}_{--}(t,\tau ) &=&\langle \mathcal{B}|\mathbb{G}_{t+\tau
,0}^{(-)\dag }\mathbb{G}_{t+\tau ,t}^{(+)}\mathbb{G}_{t,0}^{(-)}|\mathcal{B}%
\rangle , \\
\mathrm{E}_{+-}(t,\tau ) &=&\langle \mathcal{B}|\mathbb{G}_{t,0}^{(+)\dag }%
\mathbb{G}_{t+\tau ,t}^{(-)\dag }\mathbb{G}_{t+\tau ,t}^{(+)}\mathbb{G}%
_{t,0}^{(-)}|\mathcal{B}\rangle , \\
\mathrm{E}_{-+}(t,\tau ) &=&\langle \mathcal{B}|\mathbb{G}_{t+\tau
,0}^{(-)\dag }\mathbb{G}_{t+\tau ,0}^{(+)}|\mathcal{B}\rangle .
\end{eqnarray}%
We notice that the last expression corresponds to the standard LE. In fact, $%
\mathrm{E}_{-+}(t,\tau )=\mathrm{E}(t+\tau ).$

From the analysis of Ref.~\cite{PLA_2026} we concluded that, given a complex
system, the difference $|\mathrm{E}_{++}(t,\tau )-\mathrm{E}_{--}(t,\tau )|$
quantify intrinsic non-commuting quantum properties of the dynamics, which
can be read as departures with respect to a classical noise representation.
Furthermore, $|\mathrm{E}_{+-}(t,\tau )-\mathrm{E}(\tau )\mathrm{E}^{\ast
}(t)|$ and $|\mathrm{E}_{+-}(t,\tau )-\mathrm{E}(\tau -t)|$ can be used to
measure departures from a white-noise limit and an infinite time correlation
limit respectively.

\subsection{Open quantum system dynamics}

The present studies rely on a unitary microscopic description of an open
quantum system. We consider a two-level system (qubit) interacting with its
environment. As usual, in a unitary modeling, the total Hamiltonian $H_{T}$
is split as 
\end{subequations}
\begin{equation}
H_{T}=H_{s}+H_{e}+H_{I},  \label{Htotal}
\end{equation}%
where $H_{s}$ and $H_{e}$ are the system and environment Hamiltonians
respectively, while $H_{I}$ introduces their mutual coupling. We assume that
the Hamiltonian is time-independent and consider a dephasing interaction,
fulfilling the condition%
\begin{equation}
\lbrack H_{s},H_{I}]=0.  \label{Conmutas}
\end{equation}%
Denoting with $|s\rangle $, $s=\pm $, the eigenvectors of $H_{s}$, we write%
\begin{equation}
H_{s}=\sum_{s=\pm }\varepsilon _{s}|s\rangle \langle s|,\ \ \ \ \
H_{I}=\sum_{s=\pm }|s\rangle \langle s|\otimes \delta H_{e}^{(s)}.
\label{Hint}
\end{equation}%
Here, $\varepsilon _{\pm }$ are the system eigenenergies and $\delta
H_{e}^{(\pm )}$ are two different environment operators that determine $%
H_{I}.$

Given the dephasing character of the interaction, the bipartite propagator $%
\mathbb{G}_{t,t_{0}}$ splits in two branches, each one associated with a
system eigenstate. In the \textit{Schr\"{o}dinger picture} it reads%
\begin{equation}
\mathbb{G}_{t,t_{0}}=\exp [-i(t-t_{0})H_{T}]=\sum_{s=\pm }|s\rangle \langle
s|\otimes \mathbb{G}_{t,t_{0}}^{(s)},  \label{GeneChoringer}
\end{equation}%
where the effective environment propagators are%
\begin{equation}
\mathbb{G}_{t,t_{0}}^{(\pm )}=\exp [-i(t-t_{0})H_{e}^{(\pm )}],
\label{PropaBathPM_Schr}
\end{equation}%
with 
\begin{equation}
H_{e}^{(\pm )}\equiv \varepsilon _{\pm }+H_{e}+\delta H_{e}^{(\pm )}.
\end{equation}%
Thus, depending on the system state $|\pm \rangle $ the environment evolves
with the propagators generated by the Hamiltonians $H_{e}^{(\pm )}.$ The
dynamics can also be described in an \textit{interaction picture} with
respect to the bare Hamiltonian $H_{s}+H_{e}.$ In this case, the bipartite
propagator, for simplicity also denoted as $\mathbb{G}_{t,t_{0}},$ is%
\begin{equation}
\mathbb{G}_{t,t_{0}}=\left\lceil \exp -i\int_{t_{0}}^{t}H_{T}(t^{\prime
})dt^{\prime }\right\rceil =\sum_{s=\pm }|s\rangle \langle s|\otimes \mathbb{%
G}_{t,t_{0}}^{(s)},  \label{PropaHeisenberg}
\end{equation}%
where the symbol $\lceil \cdots \rceil $ is a time-ordering superoperator
and $H_{T}(t)=\sum_{s=\pm }|s\rangle \langle s|\otimes \delta
H_{e}^{(s)}(t). $ The environment propagators are%
\begin{equation}
\mathbb{G}_{t,t_{0}}^{(\pm )}=\left\lceil \exp -i\int_{t_{0}}^{t}\delta
H_{e}^{(\pm )}(t^{\prime })dt^{\prime }\right\rceil ,  \label{GeneIntB}
\end{equation}%
where $\delta H_{e}^{(\pm )}(t)\equiv \exp (+itH_{e})\delta H_{e}^{(\pm
)}\exp (-itH_{e}).$

With the previous definitions of environment propagators, the interpretation
of the LE [Eq.~(\ref{StandardLE})] as a measure of sensitivity to
perturbations in complex systems is straightforward. In fact, in the Schr%
\"{o}dinger picture, the LE can be read as the overlap between the states
obtained by propagation with an original Hamiltonian $H$ and with a
perturbed Hamiltonian $H+\Delta H$, where $H=H_{e}^{(+)}$ and $\Delta
H=\delta H_{e}^{(-)}-\delta H_{e}^{(+)}$. Thus, the LE measures to what
extent the perturbations introduced in the Hamiltonian induce departures in
the state evolution. A similar interpretation applies in the interaction
picture.

\subsection{System density matrix evolution \label{NonOperationalLE}}

Given the previous microscopic model, the density matrix of the system can
be obtained in an exact way. Taking bipartite pure initial conditions%
\begin{equation}
|\Psi _{0}\rangle =|\psi _{0}\rangle \otimes |\mathcal{B}\rangle
=(a|+\rangle +b|-\rangle )\otimes |\mathcal{B}\rangle ,
\label{InitialBipartito}
\end{equation}%
the system state, by tracing-out the environment degrees of freedom, $\rho
_{t}=\mathrm{Tr}_{e}[\mathbb{G}_{t,0}|\Psi _{0}\rangle \langle \Psi _{0}|%
\mathbb{G}_{t,0}^{\dagger }],$ reads%
\begin{equation}
\rho _{t}=\left( 
\begin{array}{cc}
|a|^{2} & ab^{\ast }\mathrm{E}(t) \\ 
a^{\ast }b\mathrm{E}^{\ast }(t) & |b|^{2}%
\end{array}%
\right) .  \label{RhoTe}
\end{equation}%
Here, $\mathrm{E}(t)$ is the LE as defined in Eq.~(\ref{StandardLE}), with
the propagators given by Eqs.~(\ref{PropaBathPM_Schr}) or~(\ref{GeneIntB})
depending on the representation.

Thus, the system populations $\langle \pm |\rho _{t}|\pm \rangle $ do not
evolve in time. On the other hand, the coherences $\langle \pm |\rho
_{t}|\mp \rangle $ are determined by the LE. Based on this property, in Ref.~%
\cite{tana} a connection was established between memory effects and the
physical meaning of LEs. In fact, in non-operational approaches to quantum
non-Markovianity, the presence of memory effects in dephasing dynamics can
be related with revivals in the coherence decay. Thus, a non-monotonic
behavior of the LE has a direct interpretation as a manifestation of memory
effects.

\section{GLEs and operational quantum non-Markovianity}

In this section we show that GLEs emerge naturally in the context of quantum
non-Markovianity defined in an operational way.

\subsection{Memory effects in outcomes statistics}

In an operational approach to quantum non-Markovianity, an open quantum
system is considered to be subjected to a minimal set of three consecutive
projective measurements. They are performed at times $t=0,$ $t,$ and $t+\tau
.$ The corresponding outcomes are denoted as $\{x\},$ $\{y\},$ and $\{z\}$
respectively. The system dynamics is non-Markovian if there exists at least
one set of measurement choices such that the joint probability $P(z,y,x)$
for the measurement outcomes does not fulfill a \ standard (probabilistic)
Markov condition,%
\begin{equation}
P(z,y,x)\neq P(z|y)P(y|x)P(x),  \label{P3NoMarkol}
\end{equation}%
where $P(u|v)$ denotes the conditional probability of $u$ given $v.$ This
disagreement can be quantified, for example, with a conditional past-future
correlation~\cite{budiniCPF}.

Here we focus on the case of a two-level system with dynamics defined by the
Hamiltonians~(\ref{Htotal}) and~(\ref{Hint}). The three measurements are
taken to be associated each with an arbitrary direction on the Bloch sphere.
The corresponding possible post-measurement states, which also define the
measurement projectors, are denoted as 
\begin{subequations}
\label{Directions}
\begin{eqnarray}
|n_{x}\rangle &\equiv &a_{x}|+\rangle +b_{x}|-\rangle , \\
|n_{y}\rangle &\equiv &a_{y}|+\rangle +b_{y}|-\rangle , \\
|n_{z}\rangle &\equiv &a_{z}|+\rangle +b_{z}|-\rangle .
\end{eqnarray}%
Here, we are choosing $|\pm \rangle $ to be the basis states, and we remark
that in this context $x,y,z$ do not label space directions. They label both
the order of the measurements processes ($x$ first, $y$ second, and $z$
third) as well as the two possible outcomes in each case, $x=\pm 1,\ y=\pm
1, $ and$\ z=\pm 1$. In the expressions above, by normalization, the
coefficients that define directions in Bloch sphere satisfy $%
|a_{k}|^{2}+|b_{k}|^{2}=1$ $(k=x,y,z).$ In addition, since the two possible
outcome states for each measurement form an orthonormal basis, using the
completeness relation one has $\sum_{k=\pm 1}|a_{k}|^{2}=\sum_{k=\pm
1}|b_{k}|^{2}=1$ and $\sum_{k=\pm 1}a_{k}b_{k}^{\ast }=0.$

Given the definition of the system dynamics and the measurement processes,
it is possible to formally calculate the corresponding outcome
probabilities. The explicit steps are given in the Appendix. Below we
analyze the final expressions as well as different limits of interest.

\subsubsection{Outcome probabilities}

Assuming the initial separable condition~(\ref{InitialBipartito}), the
probability $P(x)$ for the first measurement outcomes at time $t=0$ read 
\end{subequations}
\begin{equation}
P(x)=|\langle n_{x}|\psi _{0}\rangle |^{2}.
\end{equation}%
After this stage the system-environment arrangement evolves unitarily. At
time $t$ the second measurement is performed. The joint probability $P(y,x)$
of $x$- and $y$-outcomes is given by (see Appendix)%
\begin{equation}
\frac{P(y,x)}{P(x)}=|a_{y}|^{2}|a_{x}|^{2}+|b_{y}|^{2}|b_{x}|^{2}+2\mathrm{Re%
}[a_{y}^{\ast }b_{y}a_{x}b_{x}^{\ast }\mathrm{E}(t)].  \label{Pyx}
\end{equation}%
Interestingly, the time dependence of this probability\ is set by the LE
[Eq.~(\ref{StandardLE})]. Nevertheless, in contrast to non-operational
approaches to quantum non-Markovianity [see Sec.$~$(\ref{NonOperationalLE}%
)], the probability $P(y,x)=P(y|x)P(x)$ does not allow one to determine if
the outcome statistics obey or not a Markov property [Eq.$~$(\ref{P3NoMarkol}%
)].

After the second measurement, in the interval from $t$ to $t+\tau $ the
evolution is unitary, after which the third measurement is carried out. The
joint probability $P(z,y,x)$ for the three successive outcomes satisfies
(see Appendix)%
\begin{eqnarray}
\frac{P(z,y,x)}{P(x)} &=&\Big(|a_{z}|^{2}\,|a_{y}|^{2}+|b_{z}|^{2}%
\,|b_{y}|^{2}\Big)P(y|x)  \notag \\
&&+2\mathrm{Re}\Big[a_{z}^{\ast }b_{z}a_{y}b_{y}^{\ast }\mathrm{M}%
_{yx}(t,\tau )\Big],  \label{P3Myx}
\end{eqnarray}%
where $P(y|x)=P(y,x)/P(x)$ is given in Eq.$~$(\ref{Pyx}) and%
\begin{equation}
\mathrm{M}_{yx}(t,\tau )=\langle \Psi _{t}^{x}|n_{y}\rangle \,\mathbb{G}%
_{t+\tau ,t}^{(-)\dag }\mathbb{G}_{t+\tau ,t}^{(+)}\,\langle n_{y}|\Psi
_{t}^{x}\rangle .  \label{MyxDesdeMediciones}
\end{equation}%
Here, $|\Psi _{t}^{x}\rangle $ is the bipartite state at time $t$ given that
the first measurement has led to outcome $x$. Thus, $\mathrm{M}_{yx}(t,\tau
) $ is the mean value of the shifted echo operator $\mathbb{G}_{t+\tau
,t}^{(-)\dag }\mathbb{G}_{t+\tau ,t}^{(+)}$ in the (unnormalized)
environment state $\langle n_{y}|\Psi _{t}^{x}\rangle .$\ Explicitly, it can
be written as (see Appendix) 
\begin{multline}
\mathrm{M}_{yx}(t,\tau )=|a_{y}|^{2}\,|a_{x}|^{2}\,\mathrm{E}_{++}(t,\tau
)+|b_{y}|^{2}\,|b_{x}|^{2}\,\mathrm{E}_{--}(t,\tau ) \\
+a_{y}\,b_{y}^{\ast }\,a_{x}^{\ast }\,b_{x}\,\mathrm{E}_{+-}(t,\tau
)+a_{y}^{\ast }\,b_{y}\,a_{x}\,b_{x}^{\ast }\,\mathrm{E}_{-+}(t,\tau ),
\label{MxyCore}
\end{multline}%
where $\mathrm{E}_{\tilde{s}s}(t,\tau )$ $(\tilde{s}=\pm ,$ $s=\pm )$ are
defined by Eq.$~$(\ref{GeneralizedLE}). These expressions show that higher
joint outcome probabilities cannot be characterized only in terms of
standard LEs. Instead, in this operational approach the GLEs appear as the
quantities that determine the presence or absence of memory effects. This is
one of the main results of this section. Its features become more apparent
after analyzing a Markovian limit.

\subsubsection{Markovian limit}

A Markov limit is achieved when%
\begin{equation}
P(z,y,x)\overset{M}{=}P(z|y)P(y|x)P(x).  \label{MarkovCondition}
\end{equation}%
Taking into account Eq.$~$(\ref{Pyx}), which gives the form of $P(y|x),$ in
the Markovian regime $P(z|y)=P(z,y)/P(y)$ must analogously read%
\begin{equation}
\frac{P(z,y)}{P(y)}\overset{M}{=}%
(|a_{z}|^{2}|a_{y}|^{2}+|b_{z}|^{2}|b_{y}|^{2})+2\mathrm{Re}[a_{z}^{\ast
}b_{z}a_{y}b_{y}^{\ast }\mathrm{E}(\tau |t)],
\end{equation}%
where the shifted LE is%
\begin{equation}
\mathrm{E}(\tau |t)\equiv \langle \mathcal{B}|\mathbb{G}_{t+\tau
,t}^{(-)\dag }\mathbb{G}_{t+\tau ,t}^{(+)}|\mathcal{B}\rangle .
\end{equation}%
Stationary environments fulfill $\mathrm{E}(\tau |t)=\mathrm{E}(\tau ).$
Imposing the Markov structure$~$(\ref{MarkovCondition}) into Eq.$~$(\ref%
{P3Myx}), it follows that the possibility of attaining this regime, \textit{%
independently} of which measurement processes are performed, translates to a
condition on $\mathrm{M}_{yx}(t,\tau ),$ which reads%
\begin{equation}
\mathrm{M}_{yx}(t,\tau )\overset{M}{\cong }P(y|x)\mathrm{E}(\tau |t).
\end{equation}%
From Eqs.$~$(\ref{Pyx}) and$~$(\ref{MxyCore}), this condition in turn
implies that the GLEs must fulfill 
\begin{subequations}
\label{MarkovGLE}
\begin{eqnarray}
\mathrm{E}_{++}(t,\tau ) &\overset{M}{\!\cong \!}&\mathrm{E}(\tau |t), \\
\mathrm{E}_{--}(t,\tau ) &\overset{M}{\!\cong \!}&\mathrm{E}(\tau |t), \\
\mathrm{E}_{+-}(t,\tau ) &\overset{M}{\!\cong }\!&\mathrm{E}(\tau |t)\mathrm{%
E}^{\ast }(t), \\
\mathrm{E}_{-+}(t,\tau ) &\!\overset{M}{\cong }\!&\mathrm{E}(\tau |t)\mathrm{%
E}(t).
\end{eqnarray}%
Given that $\mathrm{E}_{-+}(t,\tau )=\mathrm{E}(t+\tau ),$ the last
condition implies that the LE satisfies the semigroup property 
\end{subequations}
\begin{equation}
\mathrm{E}(\tau +t)\overset{M}{\cong }\mathrm{E}(\tau |t)\mathrm{E}(t).
\label{LExpor}
\end{equation}%
The equalities$~$(\ref{MarkovGLE}) guarantee that the outcome statistics is
a Markovian one, regardless of the set of measurements chosen. They in turn
imply that the standard LE obeys a semigroup property. When the environment
is stationary, that is, $\mathrm{E}(\tau |t)=\mathrm{E}(\tau ),$ the LE must
decay in a exponential way [Eq.$~$(\ref{LExpor})]. This strong condition
does not emerge in non-operational approaches to quantum Markovianity~\cite%
{tana}. In fact, in this last context a monotonous decay (exponential or
not) of the LE is equivalent to Markovianity.

\subsection{Bidirectional system-environment information exchanges}

Memory effects could be present even when the environment plays a passive
role. More precisely, with this we mean that the system dynamics can be
equivalently obtained by coupling it with an environment whose state $\sigma
_{t}=\mathrm{Tr}_{s}[\mathbb{G}_{t,0}|\Psi _{0}\rangle \langle \Psi _{0}|%
\mathbb{G}_{t,0}^{\dagger }]$, at any time, is independent of the system
degrees of freedom. In this situation a physical environment-to-system
backflow of information cannot take place$~$\cite{Budini_2022_Entropy}.
Interestingly, the operational approach allows to detect this kind of 
\textit{casual bystander environments}$~$\cite{BIF,CBE} (CBEs). We remark
that this class of reservoir can only be achieved for particular initial
conditions or when their own dynamics involve non-unitary contributions$~$%
\cite{CBE}.

The detection scheme for this scenario is as follows. The system is
subjected to the $x$-, $y$-, and $z$-measurements. In addition, after the
intermediate $y$-measurement the system state $|n_{y}\rangle $ is changed to
a (refreshed) state $|n_{\breve{y}}\rangle $ where $\breve{y}=\pm 1$ is
drawn from an \textit{arbitrary} conditional probability $\wp (\breve{y}|x)$
and the direction in the Bloch sphere is also arbitrary. With such
procedure, if the joint probability $P(z,\breve{y},x)$ does not fulfill a
Markovian property%
\begin{equation}
P(z,\breve{y},x)\neq P(z|\breve{y})\wp (\breve{y}|x)P(x),
\end{equation}%
the environment is not of casual-bystander type. Equivalently, in this
situation it is possible to affirm that memory effects rely on bidirectional
exchanges of information between the system and its environment$~$\cite{BIF}.

\subsubsection{Outcome probabilities}

For the present model, the joint probability $P(z,\breve{y},x)$ reads (see
Appendix)%
\begin{eqnarray}
&&\frac{P(z,\breve{y},x)}{P(x)}=\{|a_{z}|^{2}|a_{\breve{y}%
}|^{2}+|b_{z}|^{2}|b_{\breve{y}}|^{2}  \label{P3CBys} \\
&&\ \ \ \ \ \ \ \ \ \ \ \ \ \ \ +2\mathrm{Re}[a_{z}^{\ast }b_{z}a_{\breve{y}%
}b_{\breve{y}}^{\ast }\mathrm{C}_{x}(t,\tau )]\}\wp (\breve{y}|x),  \notag
\end{eqnarray}%
where $\mathrm{C}_{x}(t,\tau )=\sum_{y}\mathrm{M}_{yx}(t,\tau ).$ From Eq.$~$%
(\ref{MyxDesdeMediciones}), using that $\sum_{y}|n_{y}\rangle \langle n_{y}|=%
\mathrm{I}_{s}$ is the (system) identity operator, we get%
\begin{equation}
\mathrm{C}_{x}(t,\tau )=\langle \Psi _{t}^{x}|\mathbb{G}_{t+\tau ,t}^{\dag
(-)}\mathbb{G}_{t+\tau ,t}^{(+)}|\Psi _{t}^{x}\rangle .
\end{equation}%
Thus, $\mathrm{C}_{x}(t,\tau )$ is the mean value of the shifted echo
operator $(\mathrm{I}_{s}\otimes \mathbb{G}_{t+\tau ,t}^{(-)\dag }\mathbb{G}%
_{t+\tau ,t}^{(+)})$ taken with the bipartite state at time $t$ given that
the first measurement leads to the $x$-outcome, that is, $|\Psi
_{t}^{x}\rangle .$ From Eq.$~$(\ref{MxyCore}), using the properties of the
measurement coefficients, it explicitly reads%
\begin{equation}
\mathrm{C}_{x}(t,\tau )=|a_{x}|^{2}\mathrm{E}_{++}(t,\tau )+|b_{x}|^{2}%
\mathrm{E}_{--}(t,\tau ).  \label{Cx}
\end{equation}%
Therefore, bidirectional system-environment information exchanges [Eqs.$~$(%
\ref{P3CBys}) and$~$(\ref{Cx})] are also set by the GLEs. This is the second
main result of this section. On the other hand, notice that $\mathrm{E}%
_{+-}(t,\tau )$ and $\mathrm{E}_{-+}(t,\tau )$ do not play any role in the
expression for $P(z,\breve{y},x).$

\subsubsection{Casual-bystander-environment limit}

The memory effects can equivalently be obtained by coupling the system with
a casual-bystander environment when$~$\cite{BIF}%
\begin{equation}
P(z,\breve{y},x)\overset{CBE}{=}P(z|\breve{y})\wp (\breve{y}|x)P(x).
\end{equation}%
Here, there is no \textit{a priori} condition on $P(z|\breve{y}).$ From Eq.$%
~ $(\ref{P3CBys}) the CBE condition on $P(z,\breve{y},x)$ leads to%
\begin{equation}
\mathrm{C}_{x}(t,\tau )\overset{CBE}{\cong }\mathrm{C}(t,\tau )\,.
\end{equation}%
Taking into account Eq.$~$(\ref{Cx}), for this quantity to be independent of 
$x$ one needs that%
\begin{equation}
\mathrm{E}_{++}(t,\tau )\overset{CBE}{\cong }\mathrm{E}_{--}(t,\tau ).
\label{CBCondition}
\end{equation}%
Thus, in presence of memory effects (defined in an operational way), the
previous equality between GLEs implies that the action of the environment
can be represented by a CBE.

We notice that condition$~$(\ref{CBCondition}) also emerges in a Markovian
regime. Nevertheless, here the remaining Markovian conditions in %
\eqref{MarkovGLE} are not necessary. In particular, there is no specific
condition on the time decay of the LE [Eq.$~$(\ref{LExpor})]. On the
contrary, the fulfilment of a Markovian limit guarantees that the
environment is a passive one. This result is completely consistent with the
underlying ingredients of a Born-Markov approximation$~$\cite{breuerbook},
where the state of the environment is not modified by its interaction with
the system.

\subsection{Generalized echoes when a classical noise approximation applies}

The previous results rely on a microscopic Hamiltonian description. On the
other hand, in some regimes the environment action can be well approximated
by classical noise. This case allow to read the results of Ref.~\cite%
{PLA_2026} in the present context.

In classical noise approximation, the conditional reservoir propagators [see
Eqs.~(\ref{PropaBathPM_Schr}) and (\ref{GeneIntB})] can be approximated as~%
\cite{PLA_2026}%
\begin{equation}
\mathbb{G}_{t,t_{0}}^{(\pm )}\simeq \exp \left[ -i\int_{t_{0}}^{t}\xi ^{(\pm
)}(t^{\prime })dt^{\prime }\right] ,  \label{PropaPMnoise}
\end{equation}%
where $\xi ^{(\pm )}(t^{\prime })$ are classical stochastic noises (scalar
functions). The LE [Eq.~(\ref{StandardLE})] can then be written as%
\begin{equation}
\mathrm{E}(\tau )=\left\langle \exp \left[ -i\int_{0}^{\tau }\xi (t^{\prime
})dt^{\prime }\right] \right\rangle
\end{equation}%
where $\xi (t)\equiv \xi ^{(+)}(t)-\xi ^{(-)}(t)$ and $\langle \cdots
\rangle $\ denotes an average over noise realizations. Similarly, 
\begin{subequations}
\label{GLENoise}
\begin{eqnarray}
\mathrm{E}_{++}(t,\tau )\!\!\! &=&\!\!\!\!\left\langle \exp \left[
-i\int_{t}^{t+\tau }\xi (t^{\prime })dt^{\prime }\right] \right\rangle , \\
\mathrm{E}_{--}(t,\tau )\!\!\! &=&\!\!\!\!\left\langle \exp \left[
-i\int_{t}^{t+\tau }\xi (t^{\prime })dt^{\prime }\right] \right\rangle , \\
\mathrm{E}_{+-}(t,\tau )\!\!\! &=&\!\!\!\!\left\langle \exp \left[
-i\int_{t}^{t+\tau }\xi (t^{\prime })dt^{\prime }+i\int_{0}^{t}\xi
(t^{\prime })dt^{\prime }\right] \right\rangle ,\ \ \ \ \ \ \  \\
\mathrm{E}_{-+}(t,\tau )\!\!\! &=&\!\!\!\!\left\langle \exp \left[
-i\int_{0}^{t+\tau }\xi (t^{\prime })dt^{\prime }\right] \right\rangle .
\end{eqnarray}%
Here, the GLEs fulfill the relation $\mathrm{E}_{++}(t,\tau )=\mathrm{E}%
_{++}(t,\tau )=\mathrm{E}(\tau |t).$ If in addition, the noise $\xi (t)$ is
stationary (that is, its statistics are invariant under time translations)
then the first two GLEs are equal to the standard LE: $\mathrm{E}%
_{++}(t,\tau )=\mathrm{E}_{--}(t,\tau )=\mathrm{E}(\tau ).$

By assumption the statistics of the classical noise do not depend on the
dynamics of the quantum system. Thus, this model falls within the category
of CBEs. In fact, Eq.~(\ref{CBCondition}) is fulfilled. Thus, consistently
with Ref.~\cite{PLA_2026} $|\mathrm{E}_{++}(t,\tau )-\mathrm{E}_{--}(t,\tau
)|$ quantify intrinsic non-commuting quantum properties of the environment
that cannot be represented by a classical noise.

On the other hand, the property $\mathrm{E}_{-+}(t,\tau )=\mathrm{E}(t+\tau
) $ is also valid in the noise approximation. Thus, the only GLE that cannot
be inferred from the standard LE is $\mathrm{E}_{+-}(t,\tau ).$ It is simple
to check that the conditions for Markovianity [Eq.~(\ref{MarkovGLE})] $%
\mathrm{E}_{+-}(t,\tau )\overset{M}{\cong }\mathrm{E}(\tau |t)\mathrm{E}%
^{\ast }(t),$\ and $\mathrm{E}_{-+}(t,\tau )\overset{M}{\cong }\mathrm{E}%
(\tau |t)\mathrm{E}(t),$ are satisfied when the classical noise is delta
correlated (white noise)~\cite{kampen}. On the other hand, in an infinite
correlation time limit, $\xi (t^{\prime })\cong \xi ,$ it follows $\mathrm{E}%
_{+-}(t,\tau )\cong \mathrm{E}(\tau -t).$ These features also show the
consistence between the present approach and the main conclusions of Ref.~%
\cite{PLA_2026}.

\subsection{Conditional past-future correlations}

In the measurement-based approach, departures from the Markovian regime can
be quantified with a conditional past future (CPF) correlation~\cite%
{budiniCPF}. This is defined as 
\end{subequations}
\begin{equation}
C_{pf}(t,\tau )=\sum_{z,x=\pm 1}zx[P(z,x|y)-P(z|y)P(x|y)],
\end{equation}%
where the conditional probability is $P(z,x|y)=P(z,y,x)/P(y)$ with $%
P(y)=\sum_{z,x}P(z,y,x).$ The condition $C_{pf}(t,\tau )\neq 0$ measures
memory effects for a given set of measurement processes.

In the following we will see how the GLEs are related with the conditional
past-future correlations for particular choices of measurement settings. To
avoid confusion with measurement results, an upper hat symbol is added to
indicate directions in Bloch sphere. In the present analysis, the joint
probability for the three outcomes $P(z,y,x)$\ is defined by Eq.~(\ref{P3Myx}%
). Performing the three measurements in the Bloch sphere directions $\hat{x}-%
\hat{x}-\hat{x}$, assuming that $P(x)=1/2,$ it follows%
\begin{eqnarray}
C_{pf}(t,\tau ) &=&\frac{1}{2}\mathrm{Re}[\mathrm{E}_{+-}(t,\tau )+\mathrm{E}%
_{-+}(t,\tau )]  \label{xxx} \\
&&-\frac{1}{2}\mathrm{Re}[\mathrm{E}_{++}(t,\tau )+\mathrm{E}_{--}(t,\tau )]%
\mathrm{Re}[\mathrm{E}(t)]  \notag
\end{eqnarray}%
Instead for the measurements $\hat{x}-\hat{y}-\hat{x},$%
\begin{eqnarray}
C_{pf}(t,\tau ) &=&-\frac{1}{2}\mathrm{Re}[\mathrm{E}_{+-}(t,\tau )-\mathrm{E%
}_{-+}(t,\tau )]  \label{xyx} \\
&&+\frac{1}{2}\mathrm{Im}[\mathrm{E}_{++}(t,\tau )+\mathrm{E}_{--}(t,\tau )]%
\mathrm{Im}[\mathrm{E}(t)]  \notag
\end{eqnarray}%
In both cases, when the Markov condition~(\ref{MarkovGLE}) is fulfilled, it
is simple to check that $C_{pf}(t,\tau )\overset{M}{\cong }0.$

The CPF correlation can also be defined in terms of the joint probability $%
P(z,\breve{y},x)$ [Eq.~(\ref{P3CBys})],%
\begin{equation}
\breve{C}_{pf}(t,\tau )=\sum_{z,x=\pm 1}zx[P(z,x|\breve{y})-P(z|\breve{y}%
)P(x|\breve{y})].
\end{equation}%
Here, the condition $\breve{C}_{pf}(t,\tau )\neq 0$ measures departures from
a CBE. Performing the three measurements in the Bloch sphere directions $%
\hat{z}-\hat{x}-\hat{x}$, assuming that $P(x)=1/2$ and $\wp (\breve{y}%
|x)=1/2,$ one obtains%
\begin{equation}
\breve{C}_{pf}(t,\tau )=\frac{\breve{y}}{2}\mathrm{Re}[\mathrm{E}%
_{++}(t,\tau )-\mathrm{E}_{--}(t,\tau )],  \label{zxx}
\end{equation}%
while in the directions $\hat{z}-\hat{y}-\hat{x},$ under the same
assumptions,%
\begin{equation}
\breve{C}_{pf}(t,\tau )=\frac{\breve{y}}{2}\mathrm{Im}[\mathrm{E}%
_{++}(t,\tau )-\mathrm{E}_{--}(t,\tau )].  \label{zyx}
\end{equation}%
In both cases, when condition~(\ref{CBCondition}) is approached implies that 
$\breve{C}_{pf}(t,\tau )\overset{CBE}{\cong }0.$


\section{Example}

Here we study a specific example that illustrates the main theoretical
results. Despite its simplicity, the model allows us to study in a
controlled way the presence or absence of perturbations in eigenvalues
and/or eigenstates of the environment, which provides an advantage with
respect to more complex environment descriptions~\cite{PLA_2026}.

\subsection{Hamiltonians}

The model is formed by a two-level system that interacts with an environment
[Eq.~(\ref{Hint})] consisting of a set of non-interacting spins. The
spin-dependent effective Hamiltonians for the environment [Eq.~(\ref%
{PropaBathPM_Schr})] read%
\begin{equation}
H_{e}^{(+)}=\sum_{k=1}^{N}\varepsilon _{k}^{(+)}[\cos (\Delta _{k})\sigma
_{k}^{(\hat{x})}+\sin (\Delta _{k})\sigma _{k}^{(\hat{z})}],  \label{Hplus}
\end{equation}%
and similarly%
\begin{equation}
H_{e}^{(-)}=\sum_{k=1}^{N}\varepsilon _{k}^{(-)}\sigma _{k}^{(\hat{x})}.
\label{Hminus}
\end{equation}%
With $\sigma _{k}^{(j)}$ we denote the $j$-Pauli matrix $(j=\hat{x},\hat{y},%
\hat{z})$ acting on spin $k$ of the environment $(k=1,\cdots ,N).$

For simplicity, in Eq.~(\ref{InitialBipartito}), we consider a separable
initial environment condition $|\mathcal{B}\rangle =|\mathcal{B}_{1}\rangle
\otimes \cdots \otimes |\mathcal{B}_{N}\rangle ,$ where $|\mathcal{B}%
_{k}\rangle $ is the lower eigenstate of $\sigma _{k}^{(\hat{x})}.$ We also
disregard the Hamiltonians of both the system and the environment $%
(\varepsilon _{\pm }=H_{e}=0).$ With these definitions it is simple to
realize that the two effective Hamiltonians differ in the energy of each
spin of the bath $(\varepsilon _{k}^{(+)}$ and $\varepsilon _{k}^{(-)})$
and, furthermore, each angle $\Delta _{k}$ introduces a change (rotation) in
the eigenvectors of each spin Hamiltonian. Only when $\varepsilon
_{k}^{(+)}=\varepsilon _{k}^{(-)}$ and $\Delta _{k}=0$ both dynamics become
identical. Thus, given departures from these conditions, Eqs.~(\ref{Hplus})
and~(\ref{Hminus}) provide a simple model to study intrinsic perturbations
associated with differences between forward and backward protocols.

\subsection{Echoes}

The propagators $\mathbb{G}_{t,t_{0}}^{(\pm )}$ [Eq.~(\ref{PropaBathPM_Schr}%
)] associated to $H_{e}^{(\pm )}$ can be obtained straightforwardly using
that the spins in the environment do not interact. For each spin, the
evolution can be computed with the identity ${\exp [-i\omega t(\mathbf{n}%
\cdot \mathbf{\sigma )}]}=\cos (\omega t)\mathrm{I}-i\sin (\omega t)(\mathbf{%
n}.\mathbf{\sigma )}$, where $\mathbf{n}=(n_{\hat{x}},n_{\hat{y}},n_{\hat{z}%
})$ is a unit vector indicating a direction in Bloch sphere, $\mathbf{\sigma 
}=(\sigma _{\hat{x}},\sigma _{\hat{y}},\sigma _{\hat{z}})$, and $\mathrm{I}$
denotes a two-dimensional identity matrix.

In this scenario, the standard LE from Eq.~(\ref{StandardLE}) factorizes in
the form%
\begin{equation}
\mathrm{E}(\tau )=\prod_{k=1}^{N}\mathrm{E}_{k}(\tau ),  \label{LEProduct}
\end{equation}%
where each $k$-contribution is%
\begin{equation}
\mathrm{E}_{k}(\tau )\!=\!e^{-i\tau \varepsilon _{k}^{(-)}}\![e^{+i\tau
\varepsilon _{k}^{(+)}}\!\cos ^{2}(\Delta _{k}/2)+e^{-i\tau \varepsilon
_{k}^{(+)}}\!\sin ^{2}(\Delta _{k}/2)].  \label{Echo_k}
\end{equation}%
These last two expressions imply that the LE for this model can always be
written as a classical statistical superposition of random oscillatory
terms, with frequencies and probabilities given by: 
\begin{subequations}
\label{RandomFrec}
\begin{eqnarray}
\xi _{k}^{(\pm )} &=&\varepsilon _{k}^{(+)}\pm \varepsilon _{k}^{(-)}\,, \\
P(\xi _{k}^{(\pm )}) &=&\frac{1\mp \cos (\Delta _{k})}{2}\,.
\end{eqnarray}%
That is, the frequencies $\varepsilon _{k}^{(+)}\pm \varepsilon _{k}^{(-)}$
are weighted by the probabilities $\cos ^{2}(\Delta _{k}/2)$\ and $\sin
^{2}(\Delta _{k}/2)$\ respectively. As established in Ref.~\cite{PLA_2026},
in general, departures from this \textquotedblleft classical statistical
structure\textquotedblright\ can not be observed directly in the LE, but can
be identified in the GLEs.


The GLEs defined in Eq.~(\ref{GeneralizedLE}) also factorize, taking the
form: 
\end{subequations}
\begin{equation}
\mathrm{E}_{\tilde{s}s}(t,\tau )=\prod_{k=1}^{N}\mathrm{E}_{k}^{\tilde{s}%
s}(t,\tau ),  \label{GLEExample}
\end{equation}%
where 
\begin{subequations}
\label{GLE_k}
\begin{eqnarray}
\mathrm{E}_{k}^{++}(t,\tau ) &=&\mathrm{E}_{k}(\tau )+\mathrm{C}%
_{k}^{++}(t,\tau ), \\
\mathrm{E}_{k}^{--}(t,\tau ) &=&\mathrm{E}_{k}(\tau ), \\
\mathrm{E}_{k}^{+-}(t,\tau ) &=&\mathrm{E}_{k}(\tau -t)+\mathrm{C}%
_{k}^{+-}(t,\tau ), \\
\mathrm{E}_{k}^{-+}(t,\tau ) &=&\mathrm{E}_{k}(\tau +t).
\end{eqnarray}%
While the last line is related with the general property $\mathrm{E}%
_{-+}(t,\tau )=\mathrm{E}(\tau +t)$, the simple form of the second line does
not hold in general, but is a consequence of having chosen an initial state
that is an eigenstate of $H_{e}^{(-)}.$ This choice mimics a ground state
assumption in more complex dynamics~\cite{PLA_2026}. Under this assumption
for the initial state, one has that: 
\end{subequations}
\begin{equation}
\mathrm{E}_{--}(t,\tau )=\mathrm{E}(\tau ).
\end{equation}

The extra contributions in Eqs.~(\ref{GLE_k}) are 
\begin{eqnarray}
\mathrm{C}_{k}^{++}(t,\tau ) &=&2i\sin (\varepsilon _{k}^{(+)}t)\sin
[\varepsilon _{k}^{(+)}(\tau +t)]  \notag \\
&&\times \sin (\varepsilon _{k}^{(-)}\tau )\sin ^{2}(\Delta _{k}),
\end{eqnarray}%
and similarly%
\begin{eqnarray}
\mathrm{C}_{k}^{+-}(t,\tau ) &=&2i\sin (\varepsilon
_{k}^{(+)}t)[e^{i\varepsilon _{k}^{(-)}t}\sin (\varepsilon _{k}^{(+)}\tau )]
\notag \\
&&\times \sin (\varepsilon _{k}^{(-)}\tau )\sin ^{2}(\Delta _{k}).
\end{eqnarray}

The contributions $\mathrm{C}_{k}^{++}(t,\tau )$ and $\mathrm{C}%
_{k}^{+-}(t,\tau )$\ reflect the intrinsic quantum character of the
evolution. In fact, disregarding these terms $(\mathrm{C}_{k}^{++}%
\rightarrow 0$ and $\mathrm{C}_{k}^{+-}\rightarrow 0)$ all GLEs can be
written in term of the standard LE, $\mathrm{E}_{++}(t,\tau )=\mathrm{E}%
_{--}(t,\tau )=\mathrm{E}(\tau ),$ $\mathrm{E}_{+-}(t,\tau )=\mathrm{E}(\tau
-t),$ and $\mathrm{E}_{-+}(t,\tau )=\mathrm{E}(\tau +t).$ In such a case,
the conditions of a CBE are fulfilled [Eq.~(\ref{CBCondition})].

In the following we analyze the influence on the GLEs, and system memory
effects, of changes in the eigenvalues or the eigenvectors of the
environment Hamiltonian.

\subsection{Eigenvalue perturbation}

Taking $\Delta _{k}=0,$ which implies $\mathrm{C}_{k}^{++}=0$ and $\mathrm{C}%
_{k}^{+-}=0,$ from Eq.~(\ref{LEProduct}) and~(\ref{Echo_k}) we get%
\begin{equation}
\mathrm{E}(\tau )=\exp (i\tau \omega ),\ \ \ \ \ \omega \equiv
\sum_{k=1}^{N}(\varepsilon _{k}^{(+)}-\varepsilon _{k}^{(-)}).
\end{equation}%
As expected from the statistical interpretation based on Eq.~(\ref%
{RandomFrec}) (with $\Delta _{k}=0$), in this limit the Markovian and CBE
conditions are fulfilled, respectively Eqs.~(\ref{MarkovGLE}) and~(\ref%
{CBCondition}). In fact, the partial system evolution is unitary [Eq.~(\ref%
{RhoTe})], where the frequency $\omega $ defines its (induced) Hamiltonian, $%
H_{s}=(\omega /2)\sum_{s=\pm }s|s\rangle \langle s|.$ Consistently, the
partial state of the environment does not depend on the system degrees of
freedom and is time-invariant, $\mathrm{Tr}_{s}[\mathbb{G}_{t,0}|\Psi
_{0}\rangle \langle \Psi _{0}|\mathbb{G}_{t,0}^{\dagger }]=|\mathcal{B}%
\rangle \langle \mathcal{B}|.$

\subsection{Eigenvector perturbation \label{GLEPert}}

Now we consider the limit%
\begin{equation}
\varepsilon _{k}^{(+)}=\varepsilon _{k}^{(-)}=\varepsilon _{k},\ \ \ \ \ \ \
\ \ \Delta _{k}\ll 1.  \label{Conditions}
\end{equation}%
The condition $\varepsilon _{k}^{(+)}=\varepsilon _{k}^{(-)}$ is taken for
simplifying the full analysis. The small perturbation condition $\Delta
_{k}\ll 1$ allows us to expand all GLEs as series in these angles. In
particular, Eq.~(\ref{Echo_k}) can be approximated as $\mathrm{E}_{k}(\tau
)\simeq 1-(\Delta _{k}^{2}/4)(1-e^{-2i\tau \varepsilon _{k}}).$ Thus, the LE
[Eq.~(\ref{LEProduct})] can be approximated as%
\begin{equation}
\mathrm{E}(\tau )\simeq \exp (-\gamma _{\tau }-i\varphi _{\tau })
\label{EchoBosonic}
\end{equation}%
with a decay rate given by%
\begin{equation}
\gamma _{\tau }=\sum_{k=1}^{N}\frac{\Delta _{k}^{2}}{4}[1-\cos (2\varepsilon
_{k}\tau )],  \label{Gammatime}
\end{equation}%
and a phase of the form%
\begin{equation}
\varphi _{\tau }=\sum_{k=1}^{N}\frac{\Delta _{k}^{2}}{4}\sin (2\varepsilon
_{k}\tau ).  \label{phase}
\end{equation}

Performing a similar expansion in $\Delta _{k},$ for the GLEs [Eq.~(\ref%
{GLEExample})] we get%
\begin{equation}
\mathrm{E}_{++}(t,\tau )\simeq \mathrm{E}(\tau )\exp (i\Phi _{t,\tau }^{++}),
\end{equation}%
where the extra phase is%
\begin{equation}
\Phi _{t,\tau }^{++}=2(\varphi _{t}+\varphi _{\tau }-\varphi _{t+\tau }).
\label{Phis3}
\end{equation}%
Similarly, 
\begin{equation}
\mathrm{E}_{+-}(t,\tau )\simeq \exp [-(\gamma _{\tau }+\gamma _{t}+\Gamma
_{t,\tau })]\exp (i\Phi _{t,\tau }^{+-}),  \label{Average4Gdos}
\end{equation}%
where the contribution $\Gamma _{t,\tau }$ can be written as%
\begin{equation}
\Gamma _{t,\tau }=\gamma _{t}+\gamma _{\tau }-\gamma _{t+\tau },
\label{Gama3}
\end{equation}%
and the corresponding phase is%
\begin{equation}
\Phi _{t,\tau }^{+-}=(2\varphi _{t}-\varphi _{\tau +t})+\varphi _{\tau
-t}-\varphi _{\tau }.
\end{equation}

From the above expressions it follows that the decay exponent $\gamma _{\tau
}$ of Eq.~(\ref{Gammatime}) and the phase $\varphi _{\tau }$ of Eq.~(\ref%
{phase}) set the behavior of all GLEs. If one considers a particular case
such that $\varphi _{\tau }=0~\forall ~\tau $, which in turn implies $\Phi
_{t,\tau }^{++}=\Phi _{t,\tau }^{+-}=0,$ the CBE condition $\mathrm{E}%
_{++}(t,\tau )=\mathrm{E}_{--}(t,\tau )$ [Eq.~(\ref{CBCondition})] is
fulfilled. In fact, taking $\varphi _{\tau }=0$ all GLEs can be recovered
from Eq.~(\ref{GLENoise}) by considering classical Gaussian noise with zero
mean~\cite{kampen} and where $\gamma _{\tau }=\int_{0}^{\tau
}dt_{2}\int_{0}^{\tau }dt_{1}f(|t_{2}-t_{1}|)$, where $f(|\tau |)$ is the
stationary noise correlation (see Ref.~\cite{PLA_2026}). A Markovian regime
[Eq.~(\ref{MarkovGLE})]\ is achieved in a white noise limit, $%
f(|t_{2}-t_{1}|)\approx \delta (t_{2}-t_{1}),$ which in turn implies $\Gamma
_{t,\tau }=0.$

In contrast, in the more general case with $\varphi _{\tau }\neq 0$ the GLEs
for our model cannot be interpreted as resulting from classical noise. Below
we study an specific example.

\subsubsection*{Lorentzian perturbation}

We now assume that the environment can be approximated by a continuous
distribution, $\sum_{k=1}^{N}\rightarrow \int_{0}^{\infty }d\omega ,$
jointly with the assumptions%
\begin{equation}
\varepsilon _{k}\rightarrow \frac{\omega }{2},\ \ \ \ \ \ \ \Delta
_{k}^{2}\rightarrow \frac{4A^{2}\gamma }{\pi (\omega ^{2}+\gamma ^{2})}.
\label{LorentzModel}
\end{equation}%
Thus, the environment spectra is linear in frequency while the angle
perturbation follows a Lorentzian distribution characterized by the rate $%
\gamma $ and the dimensionless amplitude $A^{2}$, which must be chosen so
that the resulting angles are small.

Under the previous assumptions, from Eq.~(\ref{Gammatime}) we get a decay
exponent%
\begin{equation}
\gamma _{\tau }=\frac{A^{2}}{2}(1-e^{-\gamma \tau }),  \label{RateLorentz}
\end{equation}%
while from Eq.~(\ref{phase}) the phase reads%
\begin{equation}
\varphi _{\tau }=\frac{A^{2}}{\pi }[\cosh (\gamma \tau )\mathrm{Shi}(\gamma
\tau )-\sinh (\gamma \tau )\mathrm{Chi}(\gamma \tau )].  \label{PhaseLorentz}
\end{equation}%
Here, $\mathrm{Shi}(x)$ and $\mathrm{Chi}(x)$ are the hyperbolic sine and
cosine integrals respectively, $\mathrm{Shi}(x)=\int_{0}^{x}dt\sinh (t)/t$
and $\mathrm{Chi}(x)=\Upsilon +\ln (x)+\int_{0}^{x}dt(\cosh (t)-1)/t.$ $%
\Upsilon $ is the Euler's constant.

In the regime $\gamma \tau \ll 1,$ it is possible to approximate%
\begin{equation}
\gamma _{\tau }\simeq \frac{\tau }{\tau _{0}},\ \ \ \ \ \ \ \ \frac{1}{\tau
_{0}}\equiv \frac{A^{2}\gamma }{2},  \label{linear}
\end{equation}%
while $\varphi _{\tau }\simeq (2\tau /\tau _{0})[1-\Upsilon +\ln (1/\gamma
\tau )]/\pi ^{2}.$ Thus, in this regime the LE decays exponentially.
Contrary to the predictions of non-operational approaches to quantum
non-Markovianity~\cite{BreuerFirst,tana}, this exponential decay of the
standard echo does not guarantee a Markovian behavior in the outcomes
statistics, nor does it imply an environment that can be classically
represented as white noise, as discussed in the following.

In Fig.~1 we plot the time dependence of the decay exponent $\gamma _{\tau }$
and the phase $\varphi _{\tau }.$ Consistently, in the short time regime $%
\gamma _{\tau }$ is characterized by a linear behavior. Furthermore, in the
inset we show the decay of the absolute value of the standard LE determined
by $\gamma _{\tau }.$ For the chosen parameter values this is essentially an
exponential decay, $\exp [-\gamma _{\tau }]\simeq \exp [-\tau /\tau _{0}].$
In fact, the saturation value of $\gamma_\tau$ is such that in the
asymptotic regime $\gamma \tau \gg 1$ the limit value of the time decay is $%
\lim_{\tau \rightarrow \infty }\exp [-\gamma _{\tau }]\simeq \exp [-(\gamma
\tau _{0})^{-1}]\approx 10^{-44}.$ 
\begin{figure}[t]
\includegraphics[bb=45 860 750 1140
1140,angle=0,width=8.8cm]{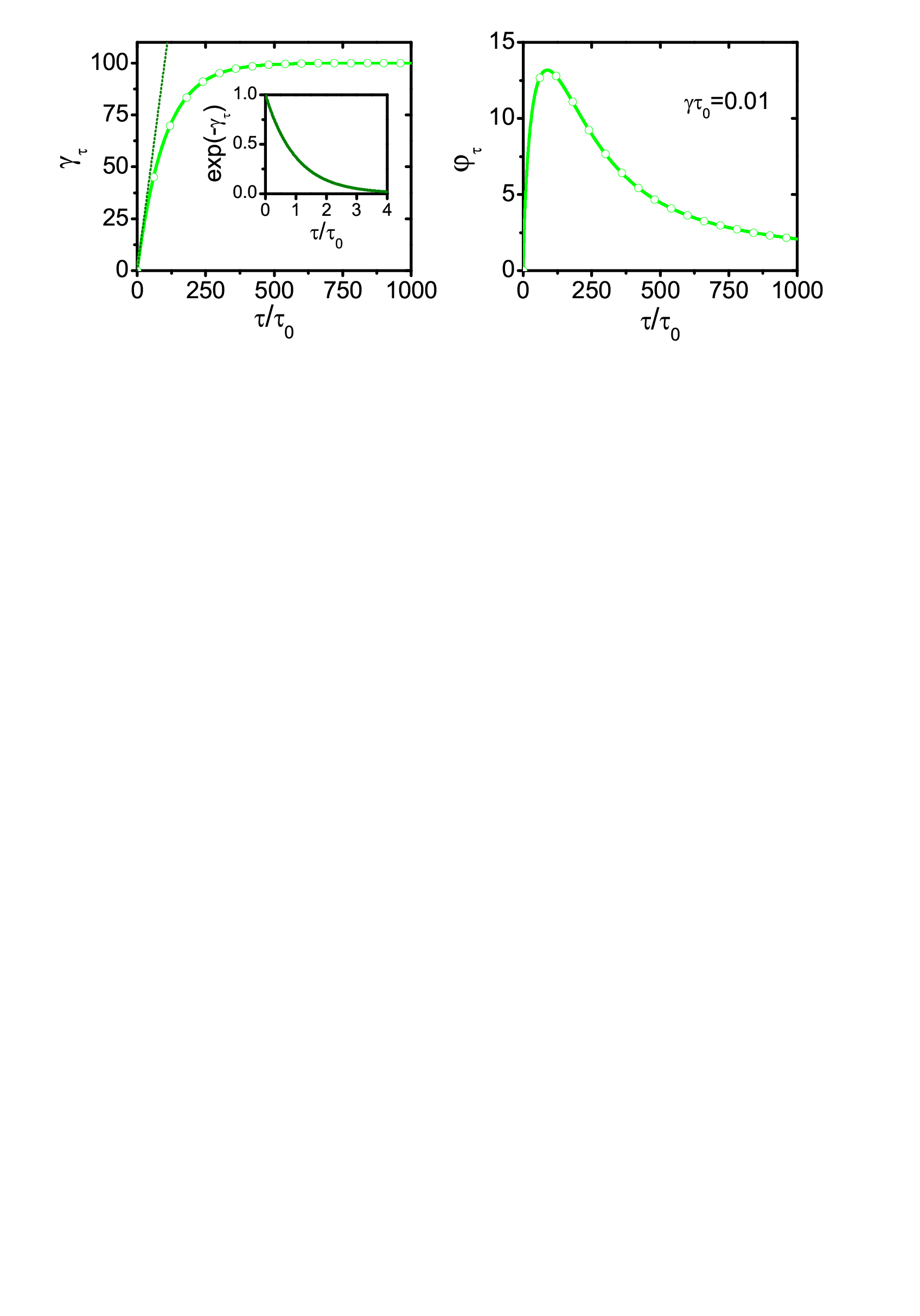}
\caption{Time dependence of the quantities that determine the echoes
associated with the Lorentzian perturbation~(\protect\ref{LorentzModel}).
Left panel: dimensionless decay exponent $\protect\gamma _{\protect\tau }$
[Eq.~(\protect\ref{RateLorentz})]. Right panel: dimensionless phase $\protect%
\varphi _{\protect\tau }$ [Eq.~(\protect\ref{PhaseLorentz})]. In the left
panel, the dotted line indicates the short-time linear regime while the
inset shows the decay $\exp (-\protect\gamma _{\protect\tau })$, closely
approximated by an exponential function. The parameters chosen are such that 
$\protect\gamma \protect\tau _{0}=10^{-2}.$}
\end{figure}

Even when the LE decay is indistinguishable from an exponential one (inset
in Fig.~1), a departure from Markovianity can be explicitly shown through
the behavior of the CPF correlations (Sec.~III D). Their explicit
expressions follow from the approximations for the GLEs obtained above. In
Fig.~2 (left panel) we plot the cases corresponding to the measurements $%
\hat{x}-\hat{x}-\hat{x}$ and $\hat{x}-\hat{y}-\hat{x},$ Eqs.~(\ref{xxx})
and~(\ref{xyx}) respectively. In both cases, clear departures from
Markovianity are observed, $C_{pf}(t,\tau )\neq 0.$

In Fig.~2 (right panel) we also plot the CPF correlations that measure
departure from a CBE-scenario. For both measurement schemes, $\hat{z}-\hat{x}%
-\hat{x}$ and $\hat{z}-\hat{y}-\hat{x},$ Eqs.~(\ref{zxx}) and~(\ref{zyx})
respectively, the feature $\breve{C}_{pf}(t,\tau )\neq 0$ implies that the
environment cannot be replaced by a classical noise.

The previous conclusions can also be revisited when both perturbation
mechanisms are taken into account, that is, when both eigenenergies and
eigenvectors of the environment spins are perturbed. If the difference $%
\varepsilon _{k}^{(+)} - \varepsilon _{k}^{(-)}$ is small, the main change
in the results is the short time behavior of the LE and its generalizations,
which instead of exponential behaviors exhibit Gaussian decays.

\section{Summary and conclusions}

We have established a solid connection between quantum non-Markovianity
defined in an operational way and complex quantum dynamics studied through
their sensibility to dynamical perturbations. This relation was established
by considering a model defined by a two-level system coupled to its
environment through a dephasing interaction. When subjecting the system to
three successive measurements, the joint outcome probability turns out to be
defined by a set of generalized Loschmidt echoes (GLEs). In analogy with the
standard LE, they are defined by state overlaps after a composition of
forward and backward propagators [Eq.~(\ref{GeneralizedLE})], but now
involving two different time intervals with durations $t$ and $\tau $.

Defining Markovianity in terms of the joint outcome probabilities, it is
possible to establish a set of conditions on the GLEs that guarantee a
Markovian behavior [Eq.~(\ref{MarkovGLE})]. In contrast to non-operational
approaches to quantum non-Markovianity, here these conditions imply that
Markovianity can only be achieved when the standard LE decays in an
exponential way, as a necessary but not sufficient condition. From the GLEs
one can also identify effects of physical system-environment information
exchanges. The presence of such exchanges implies that the evolution of the
environment state intrinsically depends on the system degrees of freedom.
Otherwise, the environment plays a passive role in the appearance of memory
effects [Eq.~(\ref{CBCondition})]. In particular, classical stochastic
noises fall within this category. 
\begin{figure}[t]
\includegraphics[bb=45 860 750 1140
1140,angle=0,width=8.8cm]{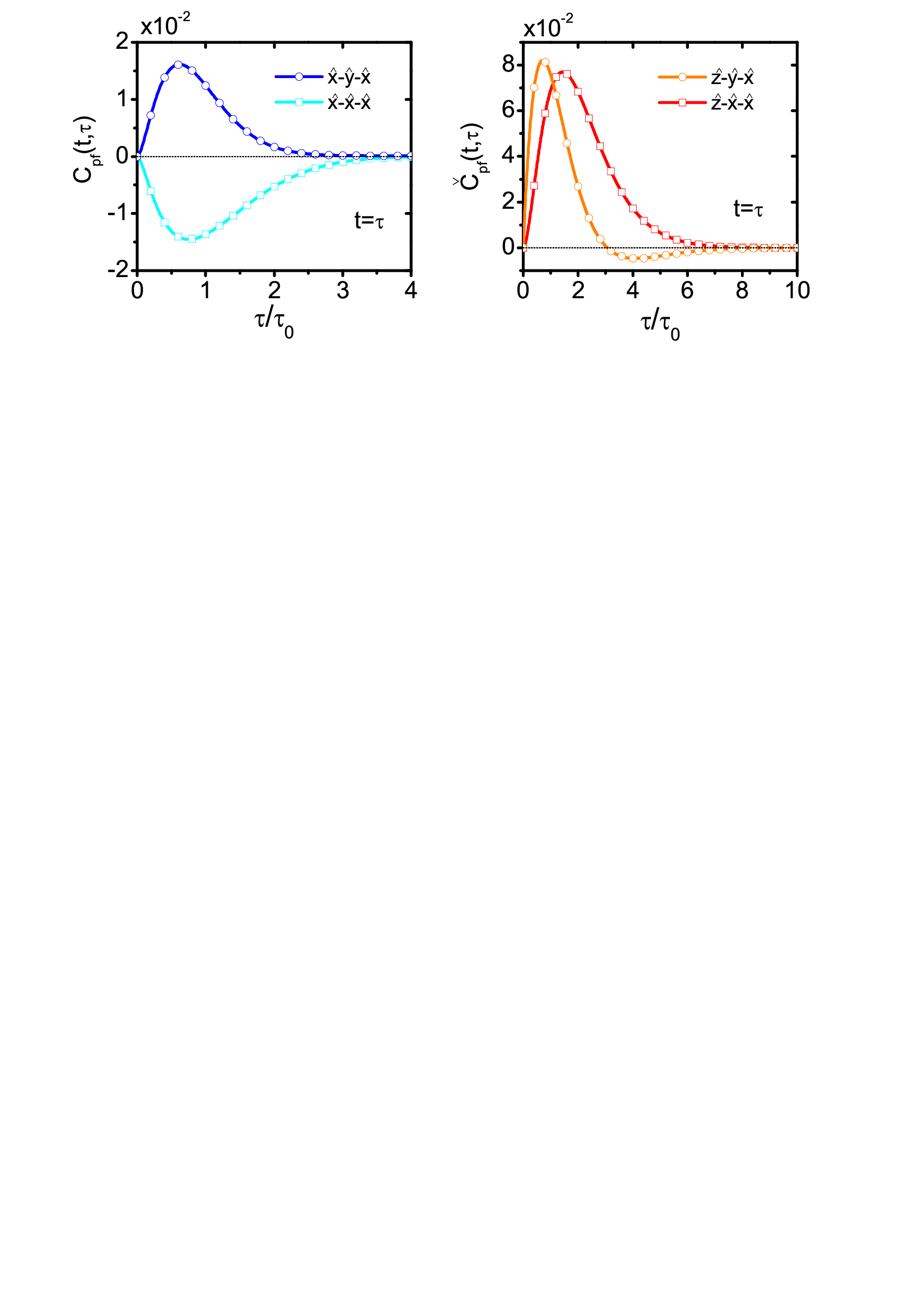}
\caption{Conditional past-future correlations resulting from the Lorentzian
model defined by Eq.~(\protect\ref{LorentzModel}). The left panel
corresponds to measurements in directions $\hat{x}-\hat{x}-\hat{x}$ and $%
\hat{x}-\hat{y}-\hat{x},$ Eqs.~(\protect\ref{xxx}) and~(\protect\ref{xyx})
respectively. The right panel corresponds to the measurements $\hat{z}-\hat{x%
}-\hat{x}$ and $\hat{z}-\hat{y}-\hat{x},$ Eqs.~(\protect\ref{zxx}) and~(%
\protect\ref{zyx}) respectively, with $\breve{y}=1.$ In both panels the
parameters are the same as in Fig.~1, $\protect\gamma \protect\tau %
_{0}=10^{-2}.$}
\end{figure}

As an example, we studied a two-level system coupled to a set of
non-interacting spins. In order to represent possible perturbations of a
complex quantum system, the Hamiltonians generating the evolution of the
environment were taken differ in either the eigenvalues or the eigenvectors
the Hamiltonian for each spin. For the chosen (reservoir) initial
conditions, the first case leads to Markovian dynamics with a passive
environment. In the second case, we considered an eigenvector perturbation,
or change of direction in Bloch sphere, which was sufficiently small to
expand all GLEs in simple analytical expressions. Considering a Lorentzian
perturbation, we concluded that even when the standard LE is characterized
by an exponential decay this feature does not guarantee that a Markovian
regime is approached in an operational perspective. These properties were
characterized in terms of conditional past-future correlations for different
measurement choices.

Our work highlights the inequivalence of different approaches to
non-Markovianity and leads to simple and natural experimental protocols to
characterize environmental time correlations and various forms of memory
effects, expanding the potential of the original Loschmidt echo and its
generalizations. Interesting open questions such as the characterization of
different time regimes of the GLEs as well as the possibility of their
explicit experimental implementation in different physical arrangements also
emerge from\ the developed approach.

\section*{Acknowledgments}

A.A.B. acknowledges funding from CONICET (Argentina). C.C. acknowledges
funding from CSIC and PEDECIBA (Uruguay).

\appendix*

\section{Measurement joint probabilities}

The following calculations can be applied to either Schr\"{o}dinger or
interaction pictures by choosing the corresponding environment propagators,
Eq.~(\ref{PropaBathPM_Schr}) or Eq.~(\ref{GeneIntB}). The measurements must
be performed in the corresponding representation. The associated projectors
are defined by Eq.~(\ref{Directions}).

For simplicity all calculations are performed with wavevectors; the
extension to mixed states is straightforward. The initial system-environment
state reads%
\begin{equation}
|\Psi _{0}\rangle =|\psi _{0}\rangle \otimes |\mathcal{B}\rangle
=(a|+\rangle +b|-\rangle )\otimes |\mathcal{B}\rangle ,  \label{InitialCI}
\end{equation}%
where $a$ and $b$ are complex normalized coefficients, $|a|^{2}+|b|^{2}=1$,
determining the initial state of the system. $|\mathcal{B}\rangle $ is the
initial environment state. After the first measurement, the initial state
becomes $|\Psi _{0}\rangle \rightarrow |\Psi _{0}^{x}\rangle ,$ where%
\begin{equation}
|\Psi _{0}^{x}\rangle =(a_{x}|+\rangle +b_{x}|-\rangle )\otimes |\mathcal{B}%
\rangle .
\end{equation}%
The probability of each option, $x=\pm 1,$ is $P(x)=|\langle n_{x}|\psi
_{0}\rangle |^{2}.$ After unitary evolution up to time $t,$ the state
becomes $|\Psi _{0}^{x}\rangle \rightarrow |\Psi _{t}^{x}\rangle ,$ with%
\begin{equation}
|\Psi _{t}^{x}\rangle =a_{x}|+\rangle \otimes |\mathcal{B}_{t}^{(+)}\rangle
+b_{x}|-\rangle \otimes |\mathcal{B}_{t}^{(-)}\rangle ,  \label{PsiX}
\end{equation}%
where%
\begin{equation}
|\mathcal{B}_{t}^{(\pm )}\rangle \equiv \mathbb{G}_{t,0}^{(\pm )}|\mathcal{B}%
\rangle .  \label{Bpm}
\end{equation}%
%
%
%

When the second measurement is performed, $|\Psi _{t}^{x}\rangle \rightarrow
|\Psi _{t}^{yx}\rangle ,$ the bipartite state becomes%
\begin{equation}
|\Psi _{t}^{yx}\rangle =(a_{y}|+\rangle +b_{y}|-\rangle )\otimes |\mathcal{B}%
_{yx}(t)\rangle .
\end{equation}%
The post-measurement bath state is%
\begin{equation}
|\mathcal{B}_{yx}(t)\rangle =\frac{\langle \hat{n}_{y}|\Psi _{t}^{x}\rangle 
}{\sqrt{|\langle \hat{n}_{y}|\Psi _{t}^{x}\rangle |^{2}}}=\frac{|\mathcal{%
\tilde{B}}_{yx}(t)\rangle }{\sqrt{P(y|x)}}.  \label{Byx(t)b}
\end{equation}%
The unnormalized contribution explicitly reads%
\begin{equation}
|\mathcal{\tilde{B}}_{yx}(t)\rangle =\langle n_{y}|\Psi _{t}^{x}\rangle
=a_{x}a_{y}^{\ast }|\mathcal{B}_{t}^{(+)}\rangle +b_{x}b_{y}^{\ast }|%
\mathcal{B}_{t}^{(-)}\rangle .  \label{ByxUnormalizado}
\end{equation}%
The normalization factor is the conditional probability of the $y$-outcome
given the previous $x$-outcome, $P(y|x)=|\langle n_{y}|\Psi _{t}^{x}\rangle
|^{2}=|\langle \mathcal{\tilde{B}}_{yx}(t)|\mathcal{\tilde{B}}%
_{yx}(t)\rangle |^{2}.$ Explicitly, it is%
\begin{equation}
P(y|x)=|a_{x}|^{2}|a_{y}|^{2}+|b_{x}|^{2}|b_{y}|^{2}+2\mathrm{Re}%
[a_{x}a_{y}^{\ast }b_{x}^{\ast }b_{y}\mathrm{E}(t)].  \label{P(y|x)}
\end{equation}%
The function $\mathrm{E}(t)$ reads%
\begin{equation}
\mathrm{E}(t)\equiv \langle \mathcal{B}_{t}^{(-)}|\mathcal{B}%
_{t}^{(+)}\rangle =\langle \mathcal{B}|\mathbb{G}_{t,0}^{(-)\dag }\mathbb{G}%
_{t,0}^{(+)}|\mathcal{B}\rangle .  \label{Coher}
\end{equation}

After the previous steps, in order to perform the scheme defined in Ref.~%
\cite{BIF}, the system state is randomly chosen\ between a set of states $%
\{\rho _{\breve{y}}\}$ $(\breve{y}=\pm 1$) with a conditional probability $%
\wp (\breve{y}|y,x).$ We take pure states $\rho _{\breve{y}}=|n_{\breve{y}%
}\rangle \langle n_{\breve{y}}|,$ with $|n_{\breve{y}}\rangle \equiv a_{%
\breve{y}}|+\rangle +b_{\breve{y}}|-\rangle .$ The bipartite state then
reads $|\Psi _{t}^{yx}\rangle \rightarrow |\Psi _{t}^{\breve{y}yx}\rangle ,$%
\begin{equation}
|\Psi _{t}^{\breve{y}yx}\rangle =(a_{\breve{y}}|+\rangle +b_{\breve{y}%
}|-\rangle )\otimes |\mathcal{B}_{yx}(t)\rangle .
\end{equation}%
After unitary evolution during a time interval $\tau ,$ this state becomes $%
|\Psi _{t}^{\breve{y}yx}\rangle \rightarrow |\Psi _{t+\tau }^{\breve{y}%
yx}\rangle ,$%
\begin{equation}
|\Psi _{t+\tau }^{\breve{y}yx}\rangle =a_{\breve{y}}|+\rangle \otimes |%
\mathcal{B}_{yx}^{(+)}(t,\tau )\rangle +b_{\breve{y}}|-\rangle \otimes |%
\mathcal{B}_{yx}^{(-)}(t,\tau )\rangle ,
\end{equation}%
where%
\begin{equation}
|\mathcal{B}_{yx}^{(\pm )}(t,\tau )\rangle \equiv \mathbb{G}_{t+\tau
,t}^{(\pm )}|\mathcal{B}_{yx}(t)\rangle .  \label{Byx(t,tau)}
\end{equation}%
The probability for the last $z$-measurement process, $P(z|\breve{y}%
,y,x)=|\langle n_{z}|\Psi _{t+\tau }^{\breve{y}yx}\rangle |^{2},$ reads%
\begin{eqnarray}
P(z|\breve{y},y,x) &=&|a_{\breve{y}}|^{2}|a_{z}|^{2}+|b_{\breve{y}%
}|^{2}|b_{z}|^{2} \\
&&+2\mathrm{Re}[a_{\breve{y}}a_{z}^{\ast }b_{\breve{y}}^{\ast }b_{z}\langle 
\mathcal{B}_{yx}^{(-)}(t,\tau )|\mathcal{B}_{yx}^{(+)}(t,\tau )\rangle ]. 
\notag
\end{eqnarray}%
Using Eqs.~(\ref{Byx(t)b}) and (\ref{Byx(t,tau)}), the scalar product is $%
\langle \mathcal{B}_{yx}^{(-)}(t,\tau )|\mathcal{B}_{yx}^{(+)}(t,\tau
)\rangle =\langle \mathcal{B}_{yx}(t)|\mathbb{G}_{t+\tau ,t}^{\dag (-)}%
\mathbb{G}_{t+\tau ,t}^{(+)}|\mathcal{B}_{yx}(t)\rangle =\langle \mathcal{%
\tilde{B}}_{yx}(t)|\mathbb{G}_{t+\tau ,t}^{\dag (-)}\mathbb{G}_{t+\tau
,t}^{(+)}|\mathcal{\tilde{B}}_{yx}(t)\rangle /P(y|x).$ Thus, we write%
\begin{eqnarray}
P(z|\breve{y},y,x) &=&|a_{\breve{y}}|^{2}|a_{z}|^{2}+|b_{\breve{y}%
}|^{2}|b_{z}|^{2}  \label{PzConditional3} \\
&&+\frac{2}{P(y|x)}\mathrm{Re}[a_{\breve{y}}a_{z}^{\ast }b_{\breve{y}}^{\ast
}b_{z}\mathrm{M}_{yx}(t,\tau )],  \notag
\end{eqnarray}%
where $\mathrm{M}_{yx}(t,\tau )\equiv \langle \mathcal{\tilde{B}}_{yx}(t)|%
\mathbb{G}_{t+\tau ,t}^{(-)\dag }\mathbb{G}_{t+\tau ,t}^{(+)}|\mathcal{%
\tilde{B}}_{yx}(t)\rangle .$ From Eq.~(\ref{ByxUnormalizado}) it can be
written as%
\begin{equation}
\mathrm{M}_{yx}(t,\tau )=\langle \Psi _{t}^{x}|n_{y}\rangle \mathbb{G}%
_{t+\tau ,t}^{(-)\dag }\mathbb{G}_{t+\tau ,t}^{(+)}\langle n_{y}|\Psi
_{t}^{x}\rangle ,  \label{CyxCorta}
\end{equation}%
which from Eq.~(\ref{PsiX}) explicitly reads%
\begin{eqnarray}
\!\!\!\!\!\!\mathrm{M}_{yx}(t,\tau )\!\!\!
&=&\!\!\!|a_{x}|^{2}|a_{y}|^{2}\langle \mathcal{B}|\mathbb{G}_{t,0}^{(+)\dag
}\mathbb{G}_{t+\tau ,t}^{(-)\dag }\mathbb{G}_{t+\tau ,t}^{(+)}\mathbb{G}%
_{t,0}^{(+)}|\mathcal{B}\rangle  \notag \\
&&\!\!\!\!\!\!\!\!+|b_{x}|^{2}|b_{y}|^{2}\langle \mathcal{B}|\mathbb{G}%
_{t,0}^{(-)\dag }\mathbb{G}_{t+\tau ,t}^{(-)\dag }\mathbb{G}_{t+\tau
,t}^{(+)}\mathbb{G}_{t,0}^{(-)}|\mathcal{B}\rangle  \notag \\
&&\!\!\!\!\!\!\!\!+a_{x}^{\ast }a_{y}b_{x}b_{y}^{\ast }\langle \mathcal{B}|%
\mathbb{G}_{t,0}^{(+)\dag }\mathbb{G}_{t+\tau ,t}^{(-)\dag }\mathbb{G}%
_{t+\tau ,t}^{(+)}\mathbb{G}_{t,0}^{(-)}|\mathcal{B}\rangle  \notag \\
&&\!\!\!\!\!\!\!\!\!+a_{x}a_{y}^{\ast }b_{x}^{\ast }b_{y}\langle \mathcal{B}|%
\mathbb{G}_{t,0}^{(-)\dag }\mathbb{G}_{t+\tau ,t}^{(-)\dag }\mathbb{G}%
_{t+\tau ,t}^{(+)}\mathbb{G}_{t,0}^{(+)}|\mathcal{B}\rangle .\ \ \ \ 
\label{Long}
\end{eqnarray}

From Bayes rule, the joint probability $P(z,\breve{y},y,x)$\ for the four
outcome events $x\rightarrow y\rightarrow \breve{y}\rightarrow z,$ reads%
\begin{equation}
P(z,\breve{y},y,x)=P(z|\breve{y},y,x)\wp (\breve{y}|y,x)P(y|x)P(x).
\end{equation}%
From Eqs.~(\ref{P(y|x)}) and~(\ref{PzConditional3}) it reads 
\begin{eqnarray}
P(z,\breve{y},y,x) &=&\{(|a_{\breve{y}}|^{2}|a_{z}|^{2}+|b_{\breve{y}%
}|^{2}|b_{z}|^{2})P(y|x) \\
&&+2\mathrm{Re}[a_{\breve{y}}a_{z}^{\ast }b_{\breve{y}}^{\ast
}b_{z}M_{yx}(t,\tau )]\}\wp (\breve{y}|y,x)P(x).  \notag
\end{eqnarray}%
Finally, the probability of interest is 
\begin{equation}
P(z,\breve{y},x)=\sum_{y}P(z,\breve{y},y,x).  \label{Pjoint3}
\end{equation}

\subsection{Deterministic scheme}

In this scheme the system state after the intermediate measurement is not
modified at all~\cite{BIF}. It corresponds to performing three successive
projective measurements. This case is recovered by taking $\wp (\breve{y}%
|y,x)=\delta _{\breve{y},y}.$ From Eq.~(\ref{Pjoint3}) we get 
\begin{eqnarray}
\!\!\!\!\!\!P(z,\breve{y},x)\! &=&\!\{(|a_{\breve{y}}|^{2}|a_{z}|^{2}+|b_{%
\breve{y}}|^{2}|b_{z}|^{2})P(\breve{y}|x)  \notag \\
&&+2\mathrm{Re}[a_{\breve{y}}a_{z}^{\ast }b_{\breve{y}}^{\ast }b_{z}\mathrm{M%
}_{\breve{y}x}(t,\tau )]\}P(x).  \label{PjointDeter}
\end{eqnarray}%
Notice that in general a Markov property is not fulfilled, $P(z,\breve{y}%
,x)\neq P(z|\breve{y})P(\breve{y}|x)P(x).$ Under the indentification $\breve{%
y}\rightarrow y,$ the previous expression for $P(z,\breve{y},x)$ recovers
Eq.~(\ref{P3Myx}).

\subsection{Random scheme}

In this scheme, after the intermediate measurement, the system state is
randomly chosen~\cite{BIF}. The corresponding conditional probability
fulfills $\wp (\breve{y}|y,x)=\wp (\breve{y}|x).$ From Eq.~(\ref{Pjoint3})
we obtain%
\begin{eqnarray}
P(z,\breve{y},x) &=&\wp (\breve{y}|x)P(x)\{|a_{\breve{y}%
}|^{2}|a_{z}|^{2}+|b_{\breve{y}}|^{2}|b_{z}|^{2}  \notag \\
&&+2\mathrm{Re}[a_{\breve{y}}a_{z}^{\ast }b_{\breve{y}}^{\ast }b_{z}\sum_{y}%
\mathrm{M}_{yx}(t,\tau )]\}.  \label{Final3}
\end{eqnarray}%
This result recovers Eq.~(\ref{P3CBys}). The dependence of $\mathrm{C}%
_{x}(t,\tau )=\sum_{y}\mathrm{M}_{yx}(t,\tau )$ on $x$-outcomes leads to
departures of $P(z,\breve{y},x)$ from a Markovian case. From Eq.~(\ref%
{CyxCorta}) and using that $\sum_{y}|n_{y}\rangle \langle n_{y}|=\mathrm{I}%
_{s},$ we get%
\begin{equation}
\sum_{y}\mathrm{M}_{yx}(t,\tau )=\langle \Psi _{t}^{x}|\mathbb{G}_{t+\tau
,t}^{\dag (-)}\mathbb{G}_{t+\tau ,t}^{(+)}|\Psi _{t}^{x}\rangle .
\end{equation}%
Thus, from Eq.~(\ref{PsiX}) it follows%
\begin{eqnarray}
\!\sum_{y}\!\mathrm{M}_{yx}(t,\tau )\! &=&\!|a_{x}|^{2}\langle \mathcal{B}|%
\mathbb{G}_{t,0}^{(+)\dag }\mathbb{G}_{t+\tau ,t}^{(-)\dag }\mathbb{G}%
_{t+\tau ,t}^{(+)}\mathbb{G}_{t,0}^{(+)}|\mathcal{B}\rangle \ \ \ \ \ \ \ \
\   \label{SumaCyx} \\
&&\!\!\!\!\!+|b_{x}|^{2}\langle \mathcal{B}|\mathbb{G}_{t,0}^{(-)\dag }%
\mathbb{G}_{t+\tau ,t}^{(-)\dag }\mathbb{G}_{t+\tau ,t}^{(+)}\mathbb{G}%
_{t,0}^{(-)}|\mathcal{B}\rangle .  \notag
\end{eqnarray}%
This result also follows from Eq.~(\ref{Long}) after taking into account
that $\sum_{y}a_{y}b_{y}^{\ast }=\sum_{y}a_{y}^{\ast }b_{y}=0$ and $%
\sum_{y}|a_{y}|^{2}=\sum_{y}|b_{y}|^{2}=1.$ These equalities are valid
independently of the chosen state $|n_{y}\rangle .$

\textit{Accidental symmetry}. If for both options $x=\pm 1$ one has that $%
|a_{x}|^{2}=|b_{x}|^{2}=1/2,$ Eqs.~(\ref{Final3}) and (\ref{SumaCyx})
immediately lead to a Markovian property $P(z,\breve{y},x)=P(z|\breve{y})\wp
(\breve{y}|x)P(x).$ This accidental symmetry is avoided by choosing the
projectors that define the first measurement (for example a measurement in
Bloch direction $\hat z$).

\end{document}